\documentstyle[aps,multicol]{revtex}
\newcommand{\rb}{\mbox{\boldmath$r$}}
\newcommand{\pb}{\mbox{\boldmath$p$}}
\begin{document}
\date{\today}
\title{What happened to the gas-liquid transition in the system of
dipolar hard spheres?}
\author{\bf Yan Levin}
\address{\it Instituto de F\'{\i}sica, Universidade Federal
do Rio Grande do Sul\\ Caixa Postal 15051, CEP 91501-970, 
Porto Alegre, RS, Brazil\\ and
Institute for Theoretical Physics, University of California,\\
Santa Barbara, CA 93106-2431, USA\\
{\small levin@if.ufrgs.br}}
\maketitle
\begin{abstract}
We explore the equilibrium properties of a system composed of dipolar
hard spheres.  A new theory based on the ideas derived from the 
work of Debye and H\"uckel, Bjerrum, and Onsager is proposed
to explain the absence of the anticipated critical
point in this system. 
\end{abstract}
\begin{multicols}{2}
\narrowtext

\centerline{{\bf PACS numbers:} 61.20.Gy; 75.50.Mm; 64.70.-p}
\bigskip

A representation of polar liquid in terms of dipolar hard spheres ($DHS$), 
$N$ rigid spheres of diameter
$a$ and dipole moment $\pb$
inside a uniform medium of dielectric constant constant $\epsilon_0$ is, 
probably, one of the most
basic statistical mechanical models.  Yet, our understanding
of this seemingly simple system is far from complete. A naive argument
based partially on intuition and partially on oversimplified
approximations suggests that as the temperature is lowered, a fluid composed 
of $DHS$ will phase separate into a coexisting
liquid and gas phases.  This conclusion 
seems to be quite intuitive, after all if 
the potential between two dipoles,
%-------------------------------------
\begin{equation}
\label{1b}
U^{dd}(\rb)=\frac{1}{\epsilon_0 r^3}\left(\pb_1\cdot\pb_2-
\frac{3(\pb_1\cdot\rb)(\pb_2\cdot\rb)}{r^2} \right) \;,
\end{equation}
%--------------------------------
is spherically averaged, one finds the familiar $ 1/r^6$ 
potential of van der 
Waals \cite{Gen70} which, of course, leads to phase separation. 
This argument, however, does not withstand the
test of computer simulations which, up to now, have failed to locate
any vestige of  phase transition\cite{Wei93,Lee93,Cai93}. 
Instead,  the simulations find that, as
the temperature is lowered, the dipolar spheres 
associate forming
polymer-like chains\cite{Lee93}.
Can the formation of chains explain the disappearance of the
liquid-gas transition?
	
To respond to this question is not easy.  In search for the
answer it is interesting to recall the mechanism of
phase separation in a different, but very related system, 
the restricted primitive model (RPM) of electrolyte\cite{Ste76}. In that
case,  ions are idealized as 
hard spheres half of which carry a positive charge, 
while the other half carry a negative charge. At low
temperatures, formation of clusters composed of positive and negative
ions is energetically favored.  First, appear dimers made
of $+-$ pairs, then  trimers  $+-+$, etc.\cite{Gil83} This  looks
very similar to the formation of chains in $DHS$, and  yet the RPM
does phase separate, while the $DHS$ do not. What is responsible for
this fundamental difference?  At first, 
one might try to appeal to purely electrostatic considerations.
Thus, it is tempting to attribute
the phase transition in the $RPM$
to the fact that by the time a cluster grows to contain  
four ions, the linear configuration becomes energetically
unfavorable, and ions tend to arrange themselves in a square.  
These compact configurations could, in principle,
provide  the nuclei for the start of condensation. It is tantalizing
to think that this is the essential difference between the
ions and the dipoles; ions energetically prefer compact clusters,
while dipoles prefer linear chains.  As appealing as this
argument might sound it is, nevertheless, incorrect.  
A careful analyses
of energies clearly demonstrates that the compact configurations
also become energetically favored for $DHS$ by the time 
the clusters grow to contain four or more dipoles \cite{Jac55}.  Thus, 
a square cluster in which the dipoles
are arranged circularly (head-to-tail) at  
forty five degrees to the lines connecting 
the centers of the nearest neighbors,
has  electrostatic  energy {\it smaller} than a linear chain of dipoles. 
The fact that the simulations
observe polymer-like chains of dipoles, instead of compact configurations
favored by electrostatics, implies that 
the entropy plays an essential role in formation of clusters.  
The energetics alone
is insufficient to explain the distinct thermodynamic 
behavior exhibited by the $RPM$ and the $DHS$.

The analogy between the $RPM$ and the $DHS$ suggests that the methods
developed to study the $RPM$ might also be applicable to the
exploration of  $DHS$.  In this respect the Debye-H\"uckel
Bjerrum theory $DHBj$ of electrolyte has proven 
particularly illuminating\cite{Ebe68,Fis93}. This
theory augments the idea of screening, introduced
by Debye and H\"uckel\cite{Deb23}, to explicitly take into account 
the formation of clusters composed of  oppositely charged ions\cite{Bje26}. 
The estimates of critical parameters
based on $DHBj$ theory are, so far, the closest to
Monte Carlo simulations\cite{Val91}.  While the idea of cluster formation
is directly applicable to  $DHS$ and has already been
exploited by various authors\cite{Jor73,Sea96,Roi96,Tav97}, 
dipoles, unlike ions, do not produce any screening.
Instead, the thermodynamic effect of dipolar motion translates into
renormalization of the effective dielectric constant of the medium.
The question that we would like to answer is whether this residual
interaction  is sufficient
to produce phase separation.  

We shall proceed in the spirit of $DHBj$ theory\cite{Fis93}.
The reduced free energy density, $f=\beta F/V$,  
of solution will be constructed as a 
sum of terms embodying the most relevant physical features of the
system, starting with the entropic  ideal gas contribution
$f^{id}=\rho \ln(\rho \Lambda^3)-\rho$.  Here $\rho=N/V$ is the density
of dipoles, $\beta=1/k_BT$, and  $\Lambda$ is the thermal wavelength.  
To obtain the
electrostatic free energy let us fix one
particle at the origin and calculate the electrostatic potential that this 
dipole feels due to the presence of other molecules.  The
electrostatic potential can be found from the solution of Laplace equation,
$\nabla^2 \phi=0$, supplemented by the appropriate boundary conditions.
We shall separate this potential into two parts, $\phi_{in}$ for
$r<a$, and $\phi_{out}$ for $r \ge a$.  Clearly  $r<a$  corresponds to
the excluded volume region into which,  due
to the hardcore repulsion,  
no other particles can penetrate. 
The boundary conditions require continuity
of the potential, $\phi_{in}(a)= \phi_{out}(a)$, 
and the displacement field, 
$\epsilon_0 \phi'_{in}(a)=\epsilon \phi'_{out}(a)$,
across the surface $r=a$. We have introduced the  renormalized  
dielectric constant of the bulk $\epsilon$,  
the expression for which can be obtained
from the Onsager's reaction field theory\cite{Ons36},
%-------------------------------------
\begin{equation}
\label{1a}
\frac{(\epsilon-\epsilon_0)(\epsilon_0+2 \epsilon)}{\epsilon}= 
4 \pi\beta p^2 \rho \;. 
\end{equation}
The Laplace equation can now be integrated
to yield the potential of the central dipole due to 
other particles.  The electrostatic {\it free energy} of
the whole system is obtained through the Debye charging 
process\cite{Fis93,Deb23}
in which all the particles in the system are charged simultaneously
from zero to their final dipolar strength,
%*************************************
\begin{equation}
\label{1}
F^{dd}=-\frac{2N p^2}{\epsilon_0 a^3} \int_0^1 \lambda {\rm d}\lambda 
\frac{\epsilon(\lambda p)-\epsilon_0}{2 \epsilon(\lambda p)+\epsilon_0};.
\end{equation}
%************************************
The integration can be done explicitly
yielding the electrostatic free
energy density~\cite{Nie71},
%*************************************
\begin{eqnarray}
\label{2}
f^{dd}&=&\frac{\beta F_{dd}}{V} 
= -\frac{1}{4\pi a^3} \left\{-2+\frac{1}{\psi(u)}+\psi(u) \right. \nonumber\\
  &+&  \left. \frac{9}{2}
\ln\left( \frac{3}{2\psi(u)+1}\right) +3 \ln\psi(u)\right\}
\end{eqnarray}
%************************************
with
%--------------------------------------
\begin{equation}
\label{3}
\epsilon(u)/\epsilon_0 \equiv \psi(u)=\frac{1}{4}\left(1+u\right)
+\frac{1}{4}\sqrt{9+2u+u^2}\;,
\end{equation}
%----------------------------------------
and $u=4\pi\rho^*/T^*$,  where we have introduced the reduced density 
$\rho^*=\rho a^3$ and the reduced temperature 
$T^*=k_B T\epsilon_0 a^3/p^2$. 	
Combining Eq.~(\ref{2}) with the entropic contribution mentioned earlier,
the total free energy density of the  system becomes 
$f=f^{id}+f^{dd}$. It is a simple matter to
see that as the temperature is lowered, this free energy violates the
thermodynamic convexity requirement, what results in a  phase 
separation
into coexisting high and low density phases~\cite{Nie71,Rus73}. 
Specifically, we find
the critical parameters to be, $\rho^*_c=0.0390807$ and $T^*_c=0.138904$. 
In principle, we could have also included the excluded volume contribution
to the total free energy, expressed through the
 free volume or Carnahan-Starling
approximation, but this would not significantly affect the location of the 
critical point\cite{Wei98}. The
fundamental conclusion of 
this Debye-H\"uckel-Onsager  theory ($DHO$)
is that the
system of dipolar hard spheres separates into a coexisting
liquid and gas phases. Can this result be trusted?  Clearly, based
on our experience with the RPM\cite{Fis93} 
this conclusion must be taken with a 
grain of salt. Just like pure $DH$, the $DHO$ theory is linear.  This
means that, although the $DHO$ is quite adequate for capturing
physics of large length scales, it fails for short distances.
In particular, the $DHO$ theory  does not take into account 
the low temperature propensity to form clusters. 
It is precisely the importance 
of these configurations which is lost in  the process 
of linearizations\cite{Ons36} 
leading
to the Onsager relation~(\ref{1a}). This conclusion is very similar to the
one reached for $RPM$\cite{Fis93}. 
A solution, in that case, has been proposed more
than seventy years ago by N. Bjerrum, who suggested that the
non-linearities, in the form of clusters, can be reintroduced into
the DH theory through the allowance of ``chemical'' association between 
particles\cite{Bje26}.  
A theory based on  Bjerrum's concept of chemical equilibrium has proven
quite successful at treating the phase separation in  $RPM$~\cite{Fis93}. 
This suggests
that the same kind of methodology might also be
be useful for studying $DHS$. 
We, thus, suppose
that at low temperatures the system consists of some free
unassociated dipoles of density $\rho_1$, as well as clusters containing
$2\le n<\infty$ hard spheres.  The density of a $n$-cluster is $\rho_n$.  
The particle conservation
requires that
%--------------------------------------
\begin{equation}
\label{4}
\rho=\sum_{n=1}^{\infty} n \rho_n \;.
\end{equation}
%----------------------------------------
Following Bjerrum,  we shall first treat clusters as non-interacting
ideal species. The interactions, therefore,  are 
restricted  to {\it unassociated} dipoles
 and their contribution to the total 
free energy density is given by Eqs.~(\ref{2}) and (\ref{3}), 
with $u=4\pi\rho^*_1/T^*$. 
In the case of RPM model, this 
approximation has proven 
to be sufficient to locate the critical point~\cite{Fis93}.   
The free energy density of a $n$-cluster 
reduces to the ideal gas form,
%--------------------------------------
\begin{equation}
\label{5}
f^{id}_n=\rho_n \ln(\rho_n \Lambda^{3n}/\zeta_n)-\rho_n \;,
\end{equation}
%----------------------------------------
where we have introduced the internal
partition function of a $n$-cluster,
%--------------------------------------
\begin{equation}
\label{6}
\zeta_n=\frac{1}{ V n!}\int \prod^n_{i=1} {\rm d^3} \rb_i  
\frac{{\rm d}\Omega_i}{4 \pi}e^{-\beta U_n^{dd}} \;.
\end{equation}
%----------------------------------------
Here $U_n^{dd}$ is the pairwise interaction potential obtained from~(\ref{1b}),
and $\Omega$'s are the relative angular orientations of dipoles forming a
cluster.  In the limit of low temperatures, where the $DHO$ predicts
the location of the critical point, 
the integrals in (\ref{6}) can be  evaluated 
for chain-like configurations to yield\cite{Jor73,Fis93},
%--------------------------------------
\begin{eqnarray}
\label{7}
\zeta_n =\left\{\frac{\pi T^{*3}}{18}\right\}^{n-1}&& \!\!\!\!\!\!
\exp\left\{\frac{n}{T^*}[\psi^{(2)}(n)-\psi^{(2)}(1)] \right. \nonumber \\
&+& \left. \frac{2}{T^*}[\psi^{(1)}(n)-\psi^{(1)}(1)]\right\} \;.
\end{eqnarray}
%----------------------------------------
where $\psi^{(1)}(n)$ and $\psi^{(2)}(n)$ are the polygamma functions
of the first and second order, respectively. The condition for
chemical equilibrium between dipoles and clusters 
is expressed through the law of mass action, $\mu_n=n\mu_1$, where
the reduced chemical potential of a species $s$ is, $\mu_s=\partial f/\partial 
\rho_s$. Substituting the total free energy density, 
$f=\sum_{n=1}^\infty  f^{id}_n+f^{dd}$, we find the distribution of cluster
densities to be
%--------------------------------------
\begin{equation}
\label{8}
\rho_n=\zeta_n \rho_1^n e^{n \mu^{ex}_1} \;,
\end{equation}
%----------------------------------------
where the excess chemical potential is defined in terms of the excess
over the ideal gas contribution, in this case, 
$\mu_s^{ex}=\partial f^{dd}/\partial \rho_s$. It is important to
note that within the Bjerrum approximation 
the excess chemical potential 
depends only on the density of {\it free} dipoles, 
and the expression~(\ref{8}) reduces to an 
infinite set of {\it decoupled}
algebraic equations.

We now make the following fundamental observation; 
since the clusters are 
ideal, their presence can only shift the critical density,  
while leaving the critical temperature unaffected\cite{Fis93}.
Thus, the critical point must {\it still} be 
located at $T^*_c=0.138904$ and must {\it still} have the
density of {\it free} dipoles $\rho^*_{1c}=0.0390807$! The distribution 
of clusters at criticality is
obtained by substituting these parameters into Eq.~(\ref{8}). In order 
for the sum in (\ref{4}) to converge, the Cauchy-Hadamard theorem
requires that 
$\Delta\equiv\lim_{n \rightarrow \infty} \rho_n^{1/n}<1$. 
Inserting the critical parameters into Eq.~(\ref{8}), we find
that at criticality $\Delta_c\approx100$ and the theorem is 
{\it strongly} violated. The critical density $\rho^*_{1c}$ lies
far outside the radius of convergence of (\ref{4}).
This means that for {\it any} 
finite density $\rho$,
the density of {\it free} dipoles remains insufficiently small to 
reach phase separation! Clearly, the argument presented above assumes
that only free particles interact while the clusters are treated
as non-interacting ideal species.  This certainly is
a very strong approximation which must be considered in more
detail,  nevertheless, we note
that a similar argument has proven to be sufficient
to locate the critical point of the RPM\cite{Fis93}. In that case, 
it was found that in the vicinity of the critical point, 
the series (\ref{4}) was very quickly convergent with most of the 
ions belonging to dipolar pairs\cite{Gil83,Fis93}. 

To explore the role
played by dipole-cluster and cluster-cluster interactions 
it is necessary to account for their 
contribution to the overall free energy.  This
is far from simple.  Some progress, however, can be made if we make
the following observation.  The electrostatic potential produced by
a rigid line of dipolar density $\pb/a$ is {\it exactly}
the same as the potential due to two fictitious  monopoles of charge $\pm p/a$
located at the line's extremities.  
This can be shown explicitly by integrating
Eq.~(\ref{1b}).  The isomorphism between line of dipoles and two
discrete monopoles suggests that for low temperatures, when
the dipolar chains are quite rigid,  the
dipole-cluster and cluster-cluster
contribution to the total free energy can be approximated 
 by the energy that is required to solvate 
$N_c=2V \sum_{n=2}^{\infty} \rho_n$ monopoles in the sea of dipoles,
and by the energy of their
mutual interaction. 
The solvation energy of an ion can be obtained
following the same method presented earlier for
calculating the dipole-dipole contribution. We find,
%*************************************
\begin{equation}
\label{9a}
F^{dc}=\frac{N_c p^2}{\epsilon_0 a^3} \int_0^1 \lambda {\rm d}\lambda 
\left[\frac{\epsilon_0}{\epsilon(\lambda p)}-1\right] \;.
\end{equation}
%************************************
Performing the integration, the reduced free energy density due to
dipole-cluster interactions is found to be
%--------------------------------------
\begin{eqnarray}
\label{9}
f^{dc}&=&\sum_{n=2}^{\infty}\frac{\rho_n}{4\pi \rho_1^*}\left\{
\frac{3}{2}-\frac{1}{2\psi^2(u)}+\frac{1}{\psi(u)} \right. \nonumber \\
 &-& \left.  2\psi(u)+
2\ln \psi(u) \frac{}{}\right\} \;.
\end{eqnarray}
%----------------------------------------
Finally, the cluster-cluster contribution can now be estimated as 
the energy of a plasma composed of $N_c$ ions inside a medium of dielectric
constant $\epsilon$. We find the familiar Debye-H\"uckel 
expression\cite{Fis93,Deb23},
%--------------------------------------
\begin{equation}
\label{10}
f^{cc}=\frac{-1}{4\pi a^3}\left[\ln(1+\kappa a)-\kappa a + 
\frac{(\kappa a)^2}{2} \right]\;,
\end{equation}
%---------------------------------------- 
where now 
%--------------------------------------
\begin{equation}
\label{11}
(\kappa a)^2=\frac{8\pi\sum_{n=2}^\infty \rho_n^*}{T^*\psi(u)} \;.
\end{equation}
%---------------------------------------- 
It is easy to check that at low temperatures both dipole-cluster
and cluster-cluster contributions  are
quite small, and are unlikely to modify the previous conclusion of
the absence of criticality in $DHS$.  
The exact calculation is rather difficult
to perform since the law of mass action, when the dipole-cluster
and cluster-cluster interactions are included into the  total free energy, 
reduces to an infinite
set of {\it coupled} algebraic equations. The preliminary analysis of these,
based on a variational approximation for the distribution of clusters,
does not, however, find any indication of phase separation. 
The details of these calculations will be presented elsewhere.

We conclude that the low temperature propensity to form weakly interacting
clusters absorbs most of the dipoles, preventing the 
density of free unassociated 
particles from reaching the minimum necessary for phase separation.

This work was supported in part by CNPq --- 
Conselho Nacional de
Desenvolvimento Cient{\'\i}fico e Tecnol{\'o}gico, FINEP --- Financiadora 
de Estudos e Projetos,
Brazil and by the National Science Foundation 
under Grant No. PHY94-07194.

\end{multicols}


\begin{thebibliography}{15}

\bibitem{Gen70} P.G. de Gennes and P.A. Pincus, Phys. Kondens. Mater.
{\bf 11}, 189 (1970).

\bibitem{Wei93} J.J. Weis and D. Levesque, Phys. Rev. Lett. {\bf 71},
2729, (1993).

\bibitem{Lee93} M.E. Leeuwen and B. Smit, Phys. Rev. Lett. {\bf 71},
3991, (1993).

\bibitem{Cai93} J.-M. Caillol, J. Chem. Phys, {\bf 98},
9835, (1993).

\bibitem{Ste76} G.R. Stell, K. C. Wu, and B. Larsen, Phys. Rev. Lett. {\bf 37},
1369, (1976).

\bibitem{Gil83} M.J. Gillan, Mol. Phys. {\bf 49},
421, (1983).

\bibitem{Jac55} I.S. Jacobs and C. P. Bean, Phys. Rev.  {\bf 100},
1060, (1955).

\bibitem{Ebe68} W. Ebeling, Z. Phys. Chem.(Leipzig) {\bf 238},
400, (1968).

\bibitem{Fis93} M.E. Fisher and Y. Levin, Phys. Rev. Lett. {\bf 71},
3826, (1993); Y. Levin and M.E. Fisher, Physica A {\bf 255}, 164 (1996).

\bibitem{Deb23} P.W. Debye and E. H\"uckel, Phys. Z. {\bf 24},
185, (1923).

\bibitem{Bje26} N. Bjerrum, Kgl. Dan. Vidensk. Selsk. Mat.-fys.
Medd. {\bf 7}, 1, (1926).

\bibitem{Val91} J.P. Valleau, J. Chem. Phys. {\bf 95},
584, (1991); A.Z. Panagiotopoulos, Fluid Phase Equib. {\bf 76},
92, (1993).

\bibitem{Jor73}  P.C. Jordan, Molec. Phys. {\bf 25},
961, (1973).

\bibitem{Sea96} R. P. Sear, Phys. Rev. Lett. {\bf 76},
2310, (1996).

\bibitem{Roi96} R. van Roij, Phys. Rev. Lett. {\bf 76},
3348, (1996).

\bibitem{Tav97} J.M. Taveres, M.M. Telo da Gama, and
M. A. Osipov, Phys. Rev. E {\bf 56},
R6252, (1997); M.A. Osipov, P.I.C. Teixeira, M.M. Telo da Gama,
Phys. Rev. E {\bf 54}, 2597, (1996).


\bibitem{Ons36} L. Onsager, J. Am. Chem. Soc.  {\bf 58},
1486, (1936).

\bibitem{Nie71} see also G. Nienhuis and J. M. Deutch, 
J. Chem. Phys. {\bf 55}, 4213, (1971); J. W. Sutherland, G. Nienhuis,
and J.M. Deuthch, Mol. Phys. {\bf 27}, 721, (1974).

\bibitem{Rus73} G.S. Rushbrooke, G. Stell, and J.S. H\o ye, 
Molec. Phys. {\bf 26}, 1199, (1973).

\bibitem{Wei98} V.C. Weiss and W. Schr\"oer, J. Chem. Phys. {\bf 108},
7747, (1998).



\end{thebibliography}
\end{document}